\documentclass[letter,aps,prl,twocolumn,groupedaddress]{revtex4}
\usepackage{graphicx} \usepackage{amssymb,amsmath,multirow}
\begin{document}
\title{Density functional theory  in  transition-metal  chemistry:  a
self-consistent   Hubbard U   approach}    \author{Heather    J.   Kulik}
\affiliation{Department   of   Materials   Science  and   Engineering,
Massachusetts  Institute  of  Technology,  Cambridge, MA  02139,  USA}
\author{Matteo   Cococcioni}   \affiliation{Department  of   Materials
Science  and  Engineering,   Massachusetts  Institute  of  Technology,
Cambridge,     MA     02139,     USA}     \author{Damian   A.  Scherlis}
\altaffiliation[Current   Address:   ]{Departamento   de   Qu\'{i}mica
Inorg\'{a}nica, Anal\'{i}tica y Qu\'{i}mica F\'{i}sica, Universidad de
Buenos  Aires}   \affiliation{Department  of  Materials   Science  and
Engineering,  Massachusetts  Institute  of Technology,  Cambridge,  MA
02139,   USA}  \author{Nicola   Marzari}   \affiliation{Department  of
Materials   Science  and   Engineering,  Massachusetts   Institute  of
Technology, Cambridge, MA 02139, USA}

\date{\today}

\begin{abstract}
Transition-metal centers are  the active sites for many
biological  and  inorganic  chemical reactions.  Notwithstanding  this
central  importance, density-functional  theory calculations  based on
generalized-gradient  approximations  often  fail to describe  energetics, multiplet
structures,  reaction  barriers,  and  geometries  around  the  active
sites.  We suggest  here an  alternative approach,  derived  from the
Hubbard  U  correction  to  solid-state  problems,  that  provides  an
excellent   agreement  with correlated-electron  quantum
chemistry calculations in test cases that range from the
ground state of Fe$_2$ and Fe$_2^-$ to the 
 addition-elimination of molecular hydrogen on FeO$^+$. The Hubbard
U  is determined  with a  novel self-consistent  procedure based  on a
linear-response approach.
\end{abstract}
  \maketitle 

Transition  metals are central to  our understanding of many  fundamental 
reactions, as active sites in naturally-existing  or synthetic  molecules 
that range from metalloporphyrins and oxidoreductases \cite{MMO93} to 
alkene metathesis catalysts \cite{schrock}  to light-harvesting photosynthetic 
complexes \cite{ps1nature}. Despite this relevance, most electronic-structure
approaches fail to describe consistently
or accurately transition-metal centers. Examples include
neutral and charged iron dimers \cite{leotheo},
FeO$^+$\cite{Shaik98}, 
Mn(salen) epoxidation catalysts \cite{CavJac03}, or hemeproteins \cite{Parrinello}.  

\begin{figure}[tb]           
\centering
\includegraphics[width=3.375in]{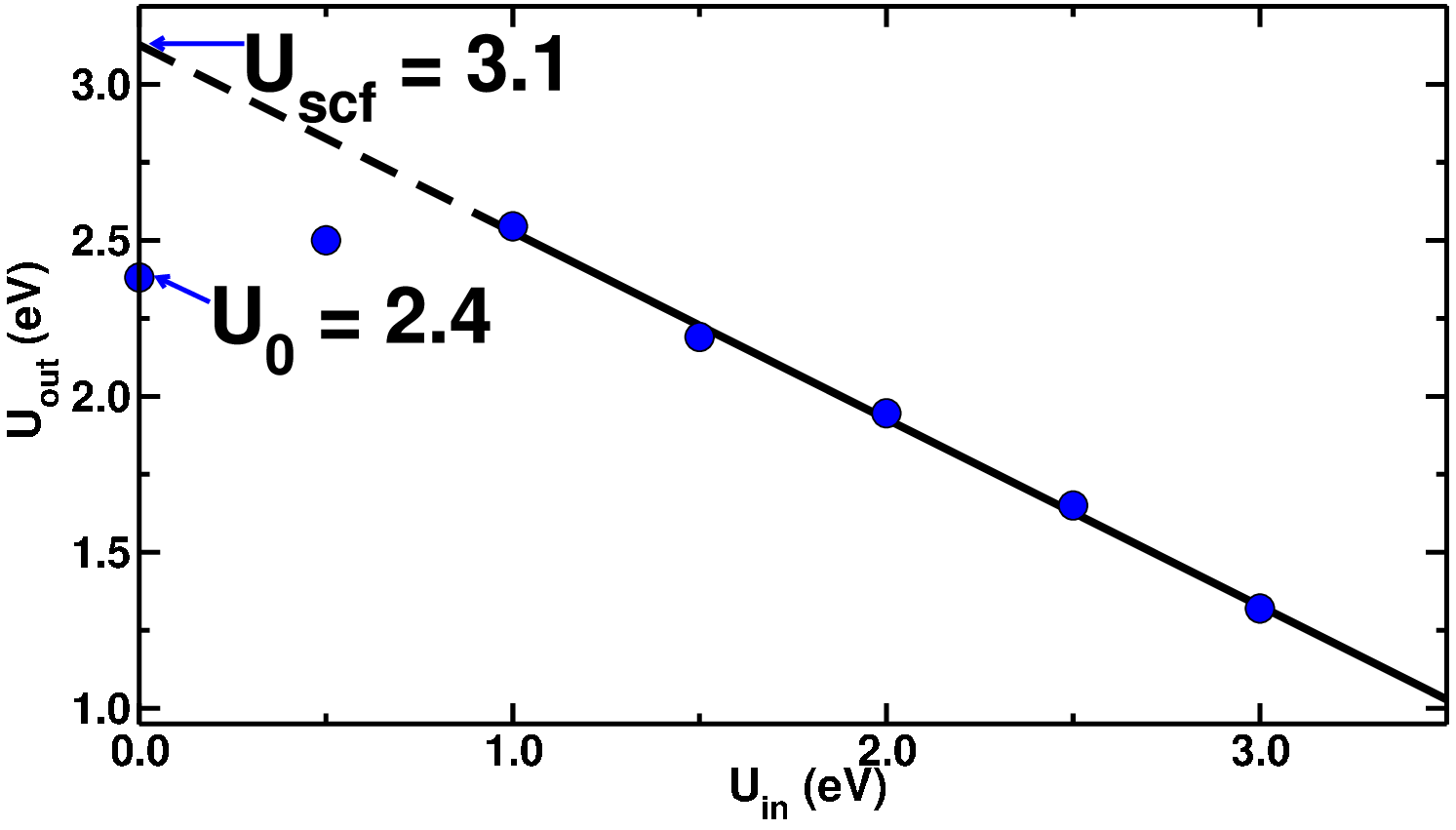}
\caption{Linear-response U$_{out}$ calculated from the GGA+U$_{in}$ ground
state  of  $^7\Delta_u$  Fe$_2$, together with the extrapolated U$_{scf}$.  U$_0$ is U$_{out}$ calculated
for U$_{in}$=0.}
\label{fig:uscf}
\end{figure}

In this Letter, we argue that generalized gradient approximations (GGA)\cite{pbe} 
augmented by a Hubbard U term\cite{anisim}, 
already very successful in the solid state \cite{LAZ95,fei},
also greatly improve single-site or few-site energies, thanks to a more
accurate description of self- and intra-atomic interactions.
Nevertheless, U is not a fitting parameter, but an intrinsic response property: 
as shown by Cococcioni and de Gironcoli \cite{GGAMC05},
U measures the spurious curvature of the GGA energy 
functional as a function of occupations, and
GGA+U largely recovers the piecewise-linear behavior 
of the exact ground-state energy.
U is determined by the difference
between the screened and bare second derivative
of the energy with respect to on-site occupations  $\lambda^I_T = 
\sum_i \lambda^{I}_i$ ($i$ is the spin-orbital, and $I$ the atomic site)\cite{GGAMC05}.
While in the original derivation U was calculated from the
GGA ground state, we argue here that U should be consistently
obtained from the GGA+U ground state itself.  This
becomes especially relevant when GGA and GGA+U 
differ qualitatively (metal vs. insulator in the solid state, different symmetry in a molecule).
To clarify our approach, we first identify in the GGA+U functional the electronic terms 
that have quadratic dependence on the occupations:
\begin{eqnarray}
\label{etot}
E_{quad}   &=&    \frac{U_{scf}}{2}\sum_{I}   \left[   \sum_{i}
\lambda^{I}_i  \left( \sum_{j} \lambda^{I}_j  - 1
\right)  \right]   \nonumber \\  
&+&  \frac{U_{in}}{2}\sum_{I}
\sum_{i} \lambda^{I}_i ( 1 -\lambda^{I}_i ).
\end{eqnarray}
The first term represents the contribution already contained in the standard GGA functional, 
modeled here as a double-counting term, 
while the second term is the customary ``+U'' correction.
Therefore, U$_{scf}$ represents the effective on-site electron-electron interaction
already present in the GGA energy  functional for the GGA+U ground state when U
is chosen to be U$_{in}$. Consistency is enforced by choosing U$_{in}$ to be equal to U$_{scf}$.
The U obtained from linear-response\cite{GGAMC05} (labeled here U$_{out}$) is 
also obtained by differentiating Eq. \ref{etot} with respect to $\lambda^I_T$:
\begin{equation}
\label{uin_uout}
U_{out}         =  
\frac{d^2E_{quad}}{d(\lambda^I_T)^2} = U_{scf} - \frac{U_{in}}{m},
\end{equation}
where $m  = 1 / \sum_i  (a^I_i)^2$ can be interpreted as an effective
degeneracy  of the orbitals  whose population  is changing  during the
perturbation (to linear order, $\delta \lambda^{I}_i = a^I_i  \delta \lambda^I_T$ 
with $\sum_i a^l_i = 1$ and $\frac{d^2}{d(\lambda^I_T)^2} = 
\sum_{ij} a^I_i  a^I_j \frac{d^2}{d\lambda^{I}_i d\lambda^{I}_j}$). 
%\footnote{The second derivatives of the energy are obtained by employing a 
%mathematical procedure identical to that described in Ref. \cite{GGAMC05}.}.
Even if in principle U$_{scf}$ depends on U$_{in}$, we
find it to be constant over a broad interval, as apparent from 
Fig. \ref{fig:uscf}:
U$_{out}$ is linear in U$_{in}$  for
the relevant range of U$_{in}$ $\sim$ U$_{scf}$. 
Thus, from few linear-response calculations for
different U$_{in}$ ground states we are able to extract the U$_{scf}$ that
should be used.

We employ this formulation in the study of the Fe$_2^-$ and Fe$_2$ dimers and
the addition-elimination reaction of molecular hydrogen on FeO$^+$:
these are paradigmatic cases of the challenges for first principles methods
 to accurately reproduce the many low-lying
multiplet potential energy surfaces associated with transition metals.  It has been
argued that spin density functional theory can describe the lowest lying state of a given
spatial and spin symmetry \cite{Ziegler,lundbarth}, but difficulties remain in obtaining
accurate multiplet splittings\cite{Filatov99}.
Our GGA or GGA+U calculations have been performed with
{\sc Quantum-ESPRESSO} \cite{spresso}; 
coupled cluster (CCSD(T)) and B3LYP calculations have been performed with Gaussian03
\cite{Gaussian}.

The iron dimer has  been investigated both theoretically \cite{leotheo,dharkes,HubSau02,Irig03}   
and experimentally \cite{mcnab,rohlf,leoexp}.  The 
experimental photoelectron spectrum of Fe$_{2}^-$ below 2 eV is remarkably
simple - there are only two prominent peaks, one at 1.0 eV and a second peak 0.53 eV above it, 
corresponding to two allowed transitions to different neutral Fe$_2$ states\cite{leoexp}. 
A recent multi-reference configuration-interaction (MRCI) study\cite{HubSau02} has assigned the three experimental electronic 
states involved as $^8\Sigma_u^-$ for Fe$_2^-$ and $^9\Sigma_g^-$ and $^7\Sigma_g^-$ for Fe$_2$; more recently, CCSD(T) has been shown to be in overall agreement\cite{Irig03}.
Importantly, these electronic states are consistent
with the experimental measurements for
the anion (fundamental frequency $\omega_0$=250$\pm$20 cm$^{-1}$ and bond length 
R$_e$=2.10$\pm$0.04 \AA\ ), and the two neutral Fe$_{2}$ states, which display
similar properties ($\omega_0$=300$\pm 15$ cm$^{-1}$ and R$_e$=2.02$\pm$0.02 \AA\ )\cite{leoexp}.

We first apply our approach to Fe$_{2}$ and Fe$_{2}^-$.  We obtain a U$_{0}$ of 2 eV (i.e. when calculated from the GGA ground state) and a U$_{scf}$ of 3 eV (since energies at different U are not directly comparable, we average U$_{0}$ and U$_{scf}$ over all states considered).  GGA+U$_{scf}$ shows a striking and consistent agreement with MRCI\cite{HubSau02} and our CCSD(T) results,
correctly identifying  both the
lowest anion state $^8\Sigma_u^-$ (3d: $\sigma_{g}^2\pi_u^4\delta_g^2\pi_g^{*,2}\delta_u^2\sigma_u^{*,1}$  
4s: $\sigma_g^2\sigma_u^{*,2}$) and the first excited 
state, $^8\Delta_g$, 0.38 eV above. The lowest, singly ionized neutral states, which differ from Fe$_{2}^-$ only by the loss of the spin down or spin up $\sigma_u^*(4s)$ orbital, are $^9\Sigma_g^{-}$ and $^7\Sigma_g^{-}$.  The $^9\Sigma_g^{-}$ $\rightarrow$ $^7\Sigma_g^{-}$  GGA+U$_{scf}$ splitting of 0.6 eV compares very well with theoretical results (MRCI and CCSD(T)) and the experimental splitting (0.53 eV) in Table \ref{table:fe2en}. 
The structure of these two states (see Table \ref{table:fe2stru}) is also consistent with 
 experimentally observed close similarity of R$_e$ and $\omega_0$ for the two neutral states and the modest decrease in R$_e$ (0.08 \AA\ ) and increase in $\omega_0$ 
($\sim$ 50 $cm^{-1}$) with respect to Fe$_2^-$ \footnote{The $\omega_e$ are obtained with a power law fit following Hollas, High Resolution Spectroscopy, Wiley, 1998.  Anharmonicity prevents direct comparison to the experimental $\omega_0$.}.     

In stark contrast with MRCI, CCSD(T) and GGA+U$_{scf}$,  GGA favors the $^8\Delta_g$  Fe$_2^-$ state (3d$^{14}$: $\sigma_{g}^2\pi_u^4\delta_g^3\pi_g^{*,2}\delta_u^2\sigma_u^{*,1}$,  
4s$^3$: $\sigma_g^2\sigma_u^{*}$) by as much as 0.9 eV relative to other methods.  Neutral states arising from single ionization of the $^8\Delta_g$ state are  $^7\Delta_u$ ($3d^{14}4s^{2}$)  and $^9\Delta_g$ ($3d^{13}4s^{3}$) which result from the loss of   $\sigma_u^*(4s)$ and $\sigma_g(3d)$ electrons, respectively.  In addition, these  two states have differing bond lengths  (R$_e$ of 1.99 and 2.26 \AA\ ) and frequencies ($\omega_e$ of 413 $cm^{-1}$ and 285 $cm^{-1}$), and thus are not compatible with experiment\cite{leotheo,leoexp}.
\begin{table} 		
\begin{tabular}{c|c|c|c|c|c|c}
\hline State & B3LYP & GGA & +U$_{0}$ &  +U$_{scf}$ & CCSD(T) & MRCI$^a$  \\ & &  & (2eV) &
(3eV)  & &\\  \hline
$^8\Sigma_g^-$& 0.00 & 0.00 & 0.00 & 0.00 & 0.00 &0.00\\ 
$^8\Delta_g$ & 0.14 & -0.52 & 0.04 & 0.38 &  0.40 & 0.45 \\
\hline
\hline
$^9\Sigma_g^-$ & 0.00  & 0.00 & 0.00 &0.00 & 0.00 & 0.00 \\  
$^7\Sigma_g^-$ &  0.34 & 0.65 & 0.66  & 0.60 &  0.55 & 0.62  \\
\hline
 $^7\Delta_u$ & 0.18  & -0.12 & 0.48 & 0.72 & 0.86 &  0.69 \\
$^9\Delta_g$ & 0.36 & 0.28 &  0.36 & 0.41 & 0.38 & 0.45  \\
\hline
\end{tabular}
\caption{Multiplet splittings (in eV) for Fe$_2^-$ and Fe$_2$ at several levels
of theory. (a) Ref. \cite{HubSau02}.}
\label{table:fe2en}
\end{table}
\begin{table} 		
\begin{tabular}{c|c|c|c|l|c}
\hline State &  GGA & GGA+U$_{scf}$ & CCSD(T) & MRCI$^a$ & Expt.$^b$ \\ 
 \hline
$^8\Sigma_g^-$ &2.20, 305 &2.20, 301 & 2.24, 276 &2.23, 272 &2.1, 250\\ 
$^8\Delta_g$ &2.07, 360&2.08, 355 &2.12, 321& 2.4,   --   &-- \\
\hline \hline $^9\Sigma_g^-$ &2.11, 339& 2.13, 335 &2.17, 296   & 2.18,299 & 2.0,300 \\ 
   $^7\Sigma_g^-$ &2.10, 335&2.12, 331& 2.16, 304 &2.17,310 &2.0,300 \\ 
\hline
   $^7\Delta_u$ &1.99, 413 &2.00, 419  & 2.00, 404 &2.25,195 &-- \\
$^9\Delta_g$ &2.26, 285 &2.26, 280 &2.28, 220&2.35,     --& -- \\
\hline
\end{tabular}
\caption{Bond lengths ( \AA\ ) and harmonic frequencies, $\omega_{e}$, ($cm^{-1}$) 
for Fe$_2^-$ and Fe$_2$, compared to experiment (here, fundamental frequencies, $\omega
_0$). (a) Ref. \cite{HubSau02}.  (b) Ref. \cite{leoexp}.}
\label{table:fe2stru}
\end{table}

Our second test case explores the potential energy surfaces of the
highly  exothermic  ($\Delta  H <  -1.6$  eV ) 
addition-elimination  reaction of molecular  hydrogen on  bare FeO$^+$.
This spin-allowed reaction
occurs with exceedingly low efficiency  (1 in every 100-1000 gas-phase collisions
results in  products), yet when  it does proceed  it is observed  to be
barrierless\cite{Armen94,Schroder97,UgaldeRev}.    This  apparent   contradiction   has been
explained by a two-state-reactivity model
\cite{Fiedler94,Shaik97,Shaik98}, wherein the steep reaction barriers along 
the spin surface of the reactants and products (sextets in both cases) preclude an efficient, exothermic 
reaction.  Instead, the reaction must occur along a shallow but excited spin surface (here, the quartet), 
and the reaction bottleneck is the coupling of the two surfaces which permits the necessary 
spin-inversion at the entrance and exit channels.    For  several  exchange-correlation functionals, (including B3LYP)\cite{Shaik97,Shaik98}, 
the reaction coordinates have failed to agree qualitatively with experiments\cite{Armen94,Schroder97,UgaldeRev}, 
higher   level   correlated-electron calculations\cite{Fiedler94,Irig99}, or
with the established paradigm of a two-state model\cite{Shaik97}.  

For the bare FeO$^+$ reactant, GGA predicts a $^6\Sigma^+$  ground state and two nearly degenerate low-lying quartet states, $^4\Delta$ and $^4\Phi$, 0.84 eV above.   GGA+U$_{scf}$ (5.5 eV) preferentially stabilizes $^4\Phi$ FeO$^+$ and yields a $^6\Sigma^+$ $\rightarrow$ $^4\Phi$ splitting of 0.54 eV in quantitative agreement with the symmetry and splitting (0.57 eV) predicted by CCSD(T). The U correction also reduces the 3$d$ character of minority spin $\pi$ molecular orbitals which dramatically improves bond lengths, harmonic frequencies,
 and anharmonicities, as shown in Table   \ref{table:feoen}.  
 \begin{table}[tb]
\centering
\begin{tabular}{c|c|c|c||c|c|c}
\hline  \multicolumn{1}{c}{ }  &  \multicolumn{3}{|c||}{$^6$FeO$^+$} &
\multicolumn{3}{c}{$^4$FeO$^+$}                \\               \hline
Method&$R_e$&$\omega_e$&$\omega_ex_e$&$R_e$&$\omega_e$&$\omega_ex_e$\\
\hline                         GGA&1.62&901&328&1.56&1038&332\\
GGA+U&1.66&749&432&1.75&612&172\\
CCSD(T)&1.66&724&434&1.70&633&188\\ \hline
\end{tabular}
\caption{Equilibrium  bond lengths, $R_e$ (\AA),  harmonic frequencies, $\omega_e$ (cm$^{-1}$),   and  anharmonicities, $\omega_ex_e$ (cm$^{-1}$)   for  the
$^6\Sigma^+$ and $^4\Phi$ states of FeO$^+$.}
\label{table:feoen}
\end{table}
\begin{figure}[tb]
\centering \includegraphics[width=3.375in]{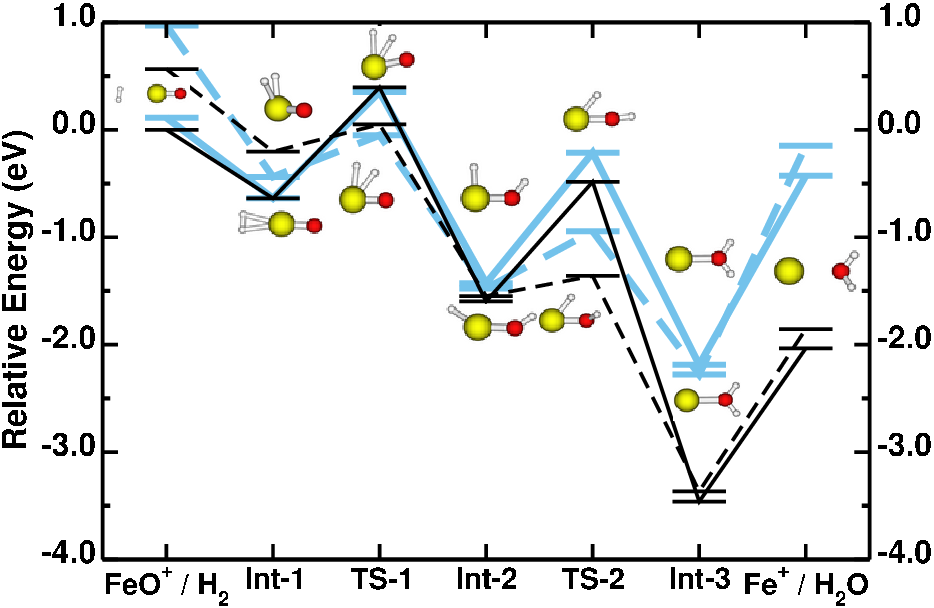}
\caption{Potential energy surface and  geometries for the FeO$^+$ + H$_{2}$
reaction  using  GGA (blue)  as compared  against
a CCSD(T)  reference  (black).   Solid  indicates  sextet  while  dashed
indicates quartet.  (Color online.)}
\label{fig:uiszero}
\centering
\end{figure}

We thus proceed to study the full sextet and quartet potential energy surfaces (PES) for this reaction.  We stress that, as is commonly found for open-shell transition-metal molecules, several low-lying  PES exist for each multiplicity and we present results for the lowest-lying symmetry of each multiplicity.  The U$_{scf}$ applied in this global PES is 5 eV, very close to the average of the U$_{scf}$ (4.93 eV) calculated for the quartet (5.02 eV)  and sextet (4.84 eV) at each stationary point; the values of U$_{0}$  are similar (quartet = 4.71 eV; sextet =  4.76 eV).  Although most states possess a U$_{scf}$ close to the global average, the few deviations will be highlighted later.

\begin{figure}[tb]
\centering \includegraphics[width=3.375in]{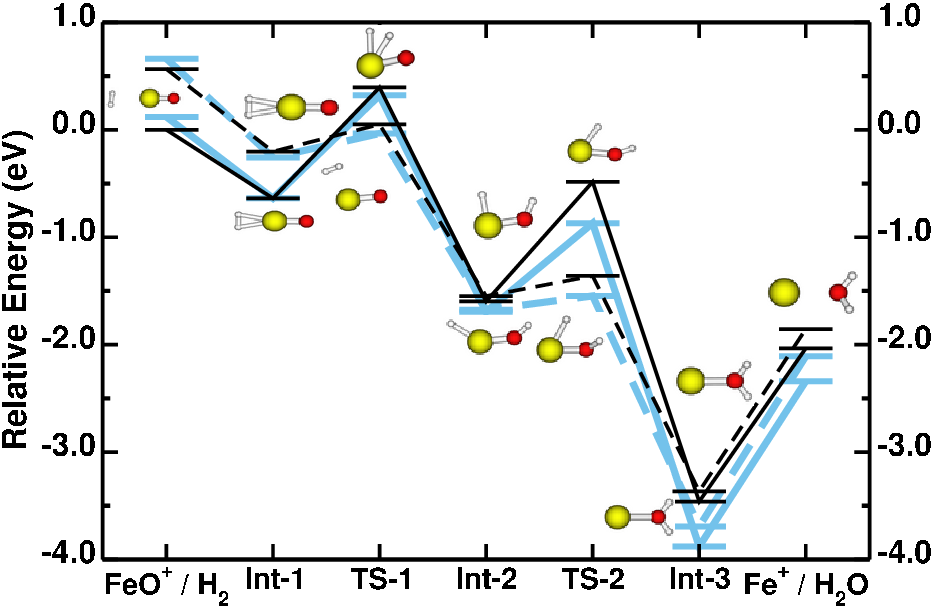}
\caption{Potential energy surface and  geometries for the FeO$^+$ + H$_{2}$
reaction  using GGA+U  (5  eV)  (blue) as  compared against  a CCSD(T)
reference  (black).   Solid indicates  sextet  while dashed  indicates
quartet, as in Fig. \ref{fig:uiszero}. (Color online.)}
\label{fig:uis5}
\end{figure}

Our  GGA results  for the intermediates  (Int) and  transition  states (TS)  along the  reaction
coordinate confirm the previously noted failures.   Aside from the overestimate of FeO$^+$ splittings, 
the most  notable
deviations  are unusually steep  barriers (0.54 eV)
along the  quartet surface, lack of  spin-crossing near the products, and a
a dramatic   underestimate  in   the
exothermicity, as  depicted in Fig.  \ref{fig:uiszero}\footnote{The PES of Figs. \ref{fig:uiszero} and \ref{fig:uis5}
have been aligned at $^6$Int-1.}.
\begin{table}
\centering
\begin{tabular}{c|ccc}
        \hline $\Delta  E_{6\rightarrow4}$ & GGA & GGA+U  & CCSD(T) \\
        \hline FeO$^+$& 0.84  & 0.54 & 0.57  \\ Int-1 & 0.20  & 0.38 &
        0.43 \\ Int-2 & -0.05 & 0.03  & 0.05 \\ Int-3 & -0.09 & 0.19(0.12) & 0.09 \\
         Fe$^+$& 0.25 & 0.22& 0.18 \\ \hline
           \end{tabular}
\centering
\caption{Multiplet  splittings (in  eV) using GGA, GGA+U (U = 5 eV except 
in parentheses, U$_{Int-3,av}$= 3.5 eV) and CCSD(T).}
\label{table:intsplits}
\end{table}
\begin{table}
\begin{tabular}{c|ccc|ccc}
\hline \multicolumn{1}{c}{  } & \multicolumn{3}{|c|}{Forward Reaction}
&  \multicolumn{3}{c}{Back Reaction} \\  \hline $\Delta  E_a$ &  GGA &
GGA+U & CCSD(T)  & GGA & GGA+U  & CCSD(T) \\ \hline TS-1$^4$  & 0.39 &
0.22 & 0.25  & 1.43 & 1.64 & 1.60  \\ TS-1$^6$ & 0.99 &  0.96 & 1.03 &
1.60 & 2.02  & 1.99 \\ TS-2$^4$ & 0.54  & 0.13 & 0.19 &  1.34 & 2.15 &
2.01 \\  TS-2$^6$ &  1.22 & 0.82(1.16)  & 1.11& 2.01  & 3.01  & 2.98 \\ 
\hline
\end{tabular}
\caption{Comparison  of GGA,  GGA+U  (U =  5  eV except in parentheses, 
U$_{4s}$ = U$_{3d}$ = 4 eV) and CCSD(T) forward
and back reaction barriers (in eV).}
\label{table:tsbarrier}
\end{table}

With GGA+U (5 eV), we obtain consistency with CCSD(T),    as   shown    in
Fig. \ref{fig:uis5}.  The reactant FeO$^+$ splitting
is  reduced, the  splitting  at Int-1 increases,
corresponding  to a  shallow quartet reaction coordinate, 
and the exothermicity and spin  crossover near the products are 
in good agreement with experiment and theoretical paradigm\cite{Shaik98}.  The quantitative accuracy of  GGA+U becomes fully evident
in  the intermediate  splittings (Table  \ref{table:intsplits}), forward  and back  reaction
barriers (Table \ref{table:tsbarrier}), and overall mean absolute errors (MAE) in multiplet 
splittings that are reduced (with respect to CCSD(T) reference)  from 0.20 eV for GGA to 0.04 eV 
for GGA+U.  Geometries are also improved: the MAE for bond lengths are reduced from 4.3 pm (GGA) to 2.2 pm (GGA+U).  \footnote{CCSD(T) geometries are from an 0.01 \AA\ interpolation 
in each degree of freedom (up to five) in total as much as 800 for a structure. 
 The states considered are reactants and intermediates for each quartet and sextet PES.}.  The GGA+U and CCSD(T) states also possess consistent orbital occupations and symmetry.

The few examples of U$_{scf}$ deviating  from  5  eV are primarily 
at  the  exit channel,  where large changes in hybridization occur.  The quartet 
Int-3  is the only case for which we obtain a low U$_{scf}$  (2  eV) 
which originates from the reduced hybridization  of Fe  $3d$ states.  We chose to recalculate the splitting with a U$_{scf,av}$ that was a local average on the Int-3 states.  With this U of 3.5 eV, we obtain a splitting of 0.12 eV, in even closer agreement with CCSD(T).  While this reduced hybridization of the 3$d$ states is unusual, we stress that it is
consistently predicted in our linear-response approach.  Along the 
sextet  surface, the iron valence occupations correspond to  $3d^64s^1$, 
and we find that the the interplay of 3$d$ 
and  4$s$ states to be critical for describing the second barrier along the sextet reaction surface.  A matrix extension of our formalism\cite{GGAMC05} considers also the response of the 4$s$ orbitals, and we obtain  U$_{4s,scf}$ =  4.0 eV and  U$_{3d,scf}$=4.0  eV around the barrier  (U$_{4s}$ is instead found to be nearly zero elsewhere).  Inclusion  of  the 4$s$  response for  both sextets Int-2 and Int-3 increases the forward reaction barrier to 1.16
eV while the backward barrier remains unchanged - in accordance with
CCSD(T).

In conclusion, we have shown how a self-consistent GGA+U approach
can provide a dramatic improvement  to the description of 
multiplet potential  energy  surfaces  for transition-metal complexes that are
otherwise poorly described by common exchange-correlation functionals,
while preserving the very favorable computational costs and scaling 
of local density-based functionals.  These improvements include
spin energetics, state symmetries, and
quantitative  description  of  complex  reaction coordinates.   
U has been treated as an intrinsic, non-empirical
property of the system considered, and never as a fitting parameter, and
it has been obtained through a self-consistent extension to the
linear-response formulation of Cococcioni and de Gironcoli\cite{GGAMC05}.
Such development will allow large-scale and accurate calculations\cite{USPP}
on transition-metal complexes, with applications in the field of
catalysis, biochemistry, and environmental science.
\begin{acknowledgments}
We  thank F. de Angelis  for pointing  out the
H$_{2}$ on FeO$^+$ reaction and S. de Gironcoli for helpful discussions on $U_{scf}$.
This work was supported by an  NSF  graduate
fellowship   and ARO-MURI  
DAAD-19-03-1-0169. Computational facilities were
provided through NSF grant DMR-0414849 and PNNL grant EMSL-UP-9597.
\end{acknowledgments}
\bibliographystyle{apsrev} 

\begin{thebibliography}{36}
\expandafter\ifx\csname natexlab\endcsname\relax\def\natexlab#1{#1}\fi
\expandafter\ifx\csname bibnamefont\endcsname\relax
  \def\bibnamefont#1{#1}\fi
\expandafter\ifx\csname bibfnamefont\endcsname\relax
  \def\bibfnamefont#1{#1}\fi
\expandafter\ifx\csname citenamefont\endcsname\relax
  \def\citenamefont#1{#1}\fi
\expandafter\ifx\csname url\endcsname\relax
  \def\url#1{\texttt{#1}}\fi
\expandafter\ifx\csname urlprefix\endcsname\relax\def\urlprefix{URL }\fi
\providecommand{\bibinfo}[2]{#2}
\providecommand{\eprint}[2][]{\url{#2}}

\bibitem[{\citenamefont{et~al.}(1993)}]{MMO93}
\bibinfo{author}{\bibfnamefont{A.~C.~Rosenzweig} \bibnamefont{et~al.}},
  \bibinfo{journal}{Nature} \textbf{\bibinfo{volume}{366}},
  \bibinfo{pages}{537} (\bibinfo{year}{1993}).

\bibitem[{\citenamefont{Schrock and Osborn}(1976)}]{schrock}
\bibinfo{author}{\bibfnamefont{R.~R.}~\bibnamefont{Schrock}} \bibnamefont{and}
  \bibinfo{author}{\bibfnamefont{J.~A.}~\bibnamefont{Osborn}},
  \bibinfo{journal}{J. Am. Chem. Soc.} \textbf{\bibinfo{volume}{98}},
  \bibinfo{pages}{2134} (\bibinfo{year}{1976}).

\bibitem[{\citenamefont{Jordan et~al.}(2001)\citenamefont{Jordan, Fromme, Witt,
  Klukas, Saenger, and Krauss}}]{ps1nature}
\bibinfo{author}{\bibfnamefont{P.~Jordan}~\bibnamefont{et al.}},
  \bibinfo{journal}{Nature} \textbf{\bibinfo{volume}{411}},
  \bibinfo{pages}{909} (\bibinfo{year}{2001}).
  
  \bibitem[{\citenamefont{et~al.}(1988)}]{leotheo}
\bibinfo{author}{\bibfnamefont{D.~G.~Leopold}~\bibnamefont{et~al.}},
  \bibinfo{journal}{J. Chem. Phys.} \textbf{\bibinfo{volume}{88}},
  \bibinfo{pages}{3780} (\bibinfo{year}{1988}).


\bibitem[{\citenamefont{Shaik and Filatov}(1998)}]{Shaik98}
\bibinfo{author}{\bibfnamefont{S.}~\bibnamefont{Shaik}} \bibnamefont{and}
  \bibinfo{author}{\bibfnamefont{M.}~\bibnamefont{Filatov}},
  \bibinfo{journal}{J. Phys. Chem. A} \textbf{\bibinfo{volume}{102}},
  \bibinfo{pages}{3835} (\bibinfo{year}{1998}).

\bibitem[{\citenamefont{Cavallo and Jacobsen}(2003)}]{CavJac03}
\bibinfo{author}{\bibfnamefont{L.}~\bibnamefont{Cavallo}} \bibnamefont{and}
  \bibinfo{author}{\bibfnamefont{H.}~\bibnamefont{Jacobsen}},
  \bibinfo{journal}{J. Phys. Chem. A} \textbf{\bibinfo{volume}{107}},
  \bibinfo{pages}{5466} (\bibinfo{year}{2003}).

\bibitem[{\citenamefont{Rovira et~al.}(2001)\citenamefont{Rovira, Schulze,
  Eichinger, Evanseck, and Parrinello}}]{Parrinello}
\bibinfo{author}{\bibfnamefont{C.~Rovira}~\bibnamefont{et~al.}},
  \bibinfo{journal}{Biophys. J.} \textbf{\bibinfo{volume}{81}},
  \bibinfo{pages}{435} (\bibinfo{year}{2001}).

\bibitem[{\citenamefont{Perdew et~al.}(1996)\citenamefont{Perdew, Burke, and
  Ernzerhof}}]{pbe}
\bibinfo{author}{\bibfnamefont{J.~P.}~\bibnamefont{Perdew}},
  \bibinfo{author}{\bibfnamefont{K.}~\bibnamefont{Burke}}, \bibnamefont{and}
  \bibinfo{author}{\bibfnamefont{M.}~\bibnamefont{Ernzerhof}},
  \bibinfo{journal}{Phys. Rev. Lett.} \textbf{\bibinfo{volume}{77}},
  \bibinfo{pages}{3865} (\bibinfo{year}{1996}).

\bibitem[{\citenamefont{Anisimov et~al.}(1991)\citenamefont{Anisimov, Zaanen,
  and Anderson}}]{anisim}
\bibinfo{author}{\bibfnamefont{V.~I.}~\bibnamefont{Anisimov}},
  \bibinfo{author}{\bibfnamefont{J.}~\bibnamefont{Zaanen}}, \bibnamefont{and}
  \bibinfo{author}{\bibfnamefont{O.~K.}~\bibnamefont{Andersen}},
  \bibinfo{journal}{Phys. Rev. B} \textbf{\bibinfo{volume}{44}}
  \bibinfo{pages}{943}(\bibinfo{year}{1991}).

\bibitem[{\citenamefont{Zhou et~al.}(2004)}]{fei}
   \bibinfo{author}{\bibfnamefont{F.~Zhou} \bibnamefont{et~al.}},
    \bibinfo{journal}{Phys. Rev. B} \textbf{\bibinfo{volume}{70}},
     \bibinfo{pages}{235121} (\bibinfo{year}{2004}).
     
\bibitem[{\citenamefont{Liechten}(1995)}]{LAZ95}
      \bibinfo{author}{\bibfnamefont{A.~I.}~\bibnamefont{Liechtenstein}}, 
       \bibinfo{author}{\bibfnamefont{V.~I.}~\bibnamefont{Anisimov}}, \bibnamefont{and}
       \bibinfo{author}{\bibfnamefont{J.}~\bibnamefont{Zaanen}}
        \bibinfo{journal}{Phys. Rev. B} \textbf{\bibinfo{volume}{52}},
        \bibinfo{pages}{R5467} (\bibinfo{year}{1995}).
        
\bibitem[{\citenamefont{Cococcioni and de~Gironcoli}(2005)}]{GGAMC05}
\bibinfo{author}{\bibfnamefont{M.}~\bibnamefont{Cococcioni}} \bibnamefont{and}
  \bibinfo{author}{\bibfnamefont{S.}~\bibnamefont{de~Gironcoli}},
  \bibinfo{journal}{Phys. Rev. B} \textbf{\bibinfo{volume}{71}},
  \bibinfo{pages}{035105} (\bibinfo{year}{2005}).

\bibitem[{\citenamefont{Ziegler}(1977)}]{Ziegler}
 \bibinfo{author}{\bibfnamefont{T.}~\bibnamefont{Ziegler}},
 \bibinfo{author}{\bibfnamefont{A.}~\bibnamefont{Rauk}} \bibnamefont{and}
 \bibinfo{author}{\bibfnamefont{E.~J.}~\bibnamefont{Baerends}},
 \bibinfo{journal}{Theor. Chim. Acta.} \textbf{\bibinfo{volume}{43}},
 \bibinfo{pages}{261} (\bibinfo{year}{1977}).  

\bibitem[{\citenamefont{GunLund}(1976)}]{lundbarth}
 \bibinfo{author}{\bibfnamefont{O.}~\bibnamefont{Gunnarsson}} \bibnamefont{and} 
 \bibinfo{author}{\bibfnamefont{B.~I.} ~\bibnamefont{Lundqvist}}, 
 \bibinfo{journal}{Phys. Rev. B} \textbf{\bibinfo{volume}{13}},
 \bibinfo{pages}{4274}(\bibinfo{year}{1976});   \bibinfo{author}{\bibfnamefont{U.}~\bibnamefont{von~Barth}}, 
 \bibinfo{journal}{Phys. Rev. A} \textbf{\bibinfo{volume}{20}},
 \bibinfo{pages}{1693}(\bibinfo{year}{1979}).

 \bibitem[{\citenamefont{FilShaik}(1999)}]{Filatov99}
Multiplets are defined by their spin
  component along the z-axis and are thus not eigenstates of
  the square of the spin operator.  Consequences and alternatives
 are discussed more fully in  \bibinfo{author}{\bibfnamefont{M.}~\bibnamefont{Filatov}} \bibnamefont{and}
 \bibinfo{author}{\bibfnamefont{S.}~\bibnamefont{Shaik}},
 \bibinfo{journal}{J. Chem. Phys.} \textbf{\bibinfo{volume}{110}},
 \bibinfo{pages}{116}(\bibinfo{year}{1999}) and references therein. 
 
  
\bibitem[{\citenamefont{et~al}()}]{spresso}
\bibinfo{author}{\bibfnamefont{S.}~\bibnamefont{Baroni}~\bibnamefont{et~al.}},
  \bibinfo{journal}{http://www.quantum-espresso.org. 
Calculations are completed in the Perdew Burke Ernzerhof (GGA) approximation\cite{pbe} using ultrasoft pseudopotentials with a plane wave cutoff of 40 Ry and density cutoff of 480 Ry;
 transition states are obtained with the nudged elastic band method\cite{Neb00}}.


\bibitem[{\citenamefont{Henkelman et~al.}(2000)\citenamefont{Henkelman,
  Uberuaga, and Jonsson}}]{Neb00}
\bibinfo{author}{\bibfnamefont{G.}~\bibnamefont{Henkelman}},
  \bibinfo{author}{\bibfnamefont{B.~P.}~\bibnamefont{Uberuaga}},
  \bibnamefont{and} \bibinfo{author}{\bibfnamefont{H.}~\bibnamefont{Jonsson}},
  \bibinfo{journal}{J. Chem. Phys.} \textbf{\bibinfo{volume}{113}},
  \bibinfo{pages}{9901} (\bibinfo{year}{2000}).

\bibitem[{\citenamefont{et~al.}(2004)}]{Gaussian}
\bibinfo{author}{\bibfnamefont{M.~J.}~\bibnamefont{Frisch} \bibnamefont{et~al.}},
  \bibinfo{journal}{Gaussian, Inc. (2004).  Hybrid 
functional B3LYP (Becke's 3-parameter exchange and Lee, 
Yang, and Parr's correlation) and CCSD(T) calculations use 
the 6-311++G(3df,3pd) basis set.}


\bibitem[{\citenamefont{Dhar and Kestner}(1988)}]{dharkes}
\bibinfo{author}{\bibfnamefont{S.}~\bibnamefont{Dhar}} \bibnamefont{and}
  \bibinfo{author}{\bibfnamefont{N.~R.}~\bibnamefont{Kestner}},
  \bibinfo{journal}{Phys. Rev. A.} \textbf{\bibinfo{volume}{38}},
  \bibinfo{pages}{1111} (\bibinfo{year}{1988}).
 
  \bibitem[{\citenamefont{Hubner and Sauer}(2002)}]{HubSau02}
\bibinfo{author}{\bibfnamefont{O.}~\bibnamefont{Hubner}} \bibnamefont{and}
  \bibinfo{author}{\bibfnamefont{J.}~\bibnamefont{Sauer}},
  \bibinfo{journal}{Chem. Phys. Lett.} \textbf{\bibinfo{volume}{358}},
  \bibinfo{pages}{442} (\bibinfo{year}{2002}) and references therein.

\bibitem[{\citenamefont{et~al.}(2003)}]{Irig03}
\bibinfo{author}{\bibfnamefont{A.}~\bibnamefont{Irigoras} \bibnamefont{et~al.}},
  \bibinfo{journal}{Chem. Phys. Lett.} \textbf{\bibinfo{volume}{376}},
  \bibinfo{pages}{310} (\bibinfo{year}{2003}).

\bibitem[{\citenamefont{McNab et~al.}(1971)\citenamefont{McNab, Micklitz, and
  Barrett}}]{mcnab}
\bibinfo{author}{\bibfnamefont{T.}~\bibnamefont{McNab}},
  \bibinfo{author}{\bibfnamefont{H.}~\bibnamefont{Micklitz}}, \bibnamefont{and}
  \bibinfo{author}{\bibfnamefont{P.}~\bibnamefont{Barrett}},
  \bibinfo{journal}{Phys. Rev. B} \textbf{\bibinfo{volume}{4}},
  \bibinfo{pages}{3787} (\bibinfo{year}{1971}).


\bibitem[{\citenamefont{Leopold and Lineberger}(1986)}]{leoexp}
\bibinfo{author}{\bibfnamefont{D.}~\bibnamefont{Leopold}} \bibnamefont{and}
  \bibinfo{author}{\bibfnamefont{W.}~\bibnamefont{Lineberger}},
  \bibinfo{journal}{J. Chem. Phys} \textbf{\bibinfo{volume}{85}},
  \bibinfo{pages}{51} (\bibinfo{year}{1986}).

\bibitem[{\citenamefont{et~al.}(1984)}]{rohlf}
\bibinfo{author}{\bibfnamefont{E.~A.}~\bibnamefont{Rohlfing} \bibnamefont{et~al.}},
  \bibinfo{journal}{J. Chem. Phys.} \textbf{\bibinfo{volume}{81}},
  \bibinfo{pages}{3846} (\bibinfo{year}{1984}).
  

\bibitem[{\citenamefont{et~al.}(1994{\natexlab{a}})}]{Armen94}
\bibinfo{author}{\bibfnamefont{D.~E.~Clemmer} \bibnamefont{et~al.}},
  \bibinfo{journal}{J. Phys. Chem.} \textbf{\bibinfo{volume}{98}},
  \bibinfo{pages}{6522} (\bibinfo{year}{1994}{\natexlab{a}}).

  
\bibitem[{\citenamefont{et~al.}(1997)}]{Schroder97}
\bibinfo{author}{\bibfnamefont{D.~Schroeder} \bibnamefont{et~al.}},
  \bibinfo{journal}{Int. J. Mass. Spec.} \textbf{\bibinfo{volume}{161}},
  \bibinfo{pages}{175} (\bibinfo{year}{1997}).

\bibitem[{\citenamefont{et~al.}(2005)}]{UgaldeRev}
\bibinfo{author}{\bibfnamefont{J.~M.~Mercero} \bibnamefont{et~al.}},
  \bibinfo{journal}{Int. J. Mass. Spec.} \textbf{\bibinfo{volume}{240}},
  \bibinfo{pages}{37} (\bibinfo{year}{2005}) and references therein.

\bibitem[{\citenamefont{et~al.}(1994{\natexlab{b}})}]{Fiedler94}
\bibinfo{author}{\bibfnamefont{A.~Fiedler} \bibnamefont{et~al.}},
  \bibinfo{journal}{J. Am. Chem. Soc.} \textbf{\bibinfo{volume}{116}},
  \bibinfo{pages}{10734} (\bibinfo{year}{1994}{\natexlab{b}}).

\bibitem[{\citenamefont{Danovich and Shaik}(1997)}]{Shaik97}
\bibinfo{author}{\bibfnamefont{D.}~\bibnamefont{Danovich}} \bibnamefont{and}
  \bibinfo{author}{\bibfnamefont{S.}~\bibnamefont{Shaik}}, \bibinfo{journal}{J.
  Am. Chem. Soc.} \textbf{\bibinfo{volume}{119}}, \bibinfo{pages}{1773}
  (\bibinfo{year}{1997}).
  
\bibitem[{\citenamefont{Irigoras et~al.}(1999)\citenamefont{Irigoras, Fowler,
  and Ugalde}}]{Irig99}
\bibinfo{author}{\bibfnamefont{A.}~\bibnamefont{Irigoras}},
  \bibinfo{author}{\bibfnamefont{J.}~\bibnamefont{Fowler}}, \bibnamefont{and}
  \bibinfo{author}{\bibfnamefont{J.}~\bibnamefont{Ugalde}},
  \bibinfo{journal}{J. Am. Chem. Soc.} \textbf{\bibinfo{volume}{121}},
  \bibinfo{pages}{8549} (\bibinfo{year}{1999}).
  
  \bibitem[{\citenamefont{Giannozzi et~al.}(2004)\citenamefont{Giannozzi,
  Angelis, and Car}}]{USPP}
\bibinfo{author}{\bibfnamefont{P.}~\bibnamefont{Giannozzi}},
  \bibinfo{author}{\bibfnamefont{F.~De} \bibnamefont{Angelis}},
  \bibnamefont{and} \bibinfo{author}{\bibfnamefont{R.}~\bibnamefont{Car}},
  \bibinfo{journal}{J. Chem. Phys.} \textbf{\bibinfo{volume}{120}},
  \bibinfo{pages}{5903} (\bibinfo{year}{2004}).
  

\end{thebibliography}

\end{document}